**Hydrolytic and lysozymic degradability of chitosan systems with heparin-mimicking pendant groups**

Giuseppe Tronci,[1,2*] Petronela Buiga,[2] Ahmed Alhilou,[2] Thuy Do,[2] Stephen J. Russell,[1] David J. Wood[2]

[1] Nonwovens Research Group, School of Design, University of Leeds, United Kingdom

[2] School of Dentistry, St. James's University Hospital, University of Leeds, United Kingdom

**Graphical abstract**

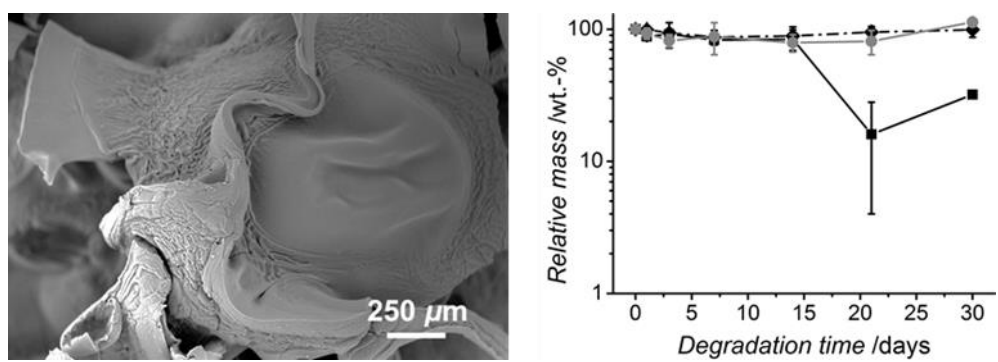

**Highlights**

- CT networks were developed with heparin-mimicking monosodium 5-sulfoisophthalate
- Tunable degradation profiles were observed in aqueous media up to 30 days
- No toxic response was observed with L929 mouse fibroblasts
- Native CT antibacterial activity was retained with *Porphyromonas gingivalis*

**Abstract**

Chitosan (CT) is an antibacterial polysaccharide that has been investigated for drug carriers, haemostats and wound dressings. For these applications, customised CT devices can often be obtained with specific experimental conditions, which can irreversibly alter native biopolymer properties and functions and lead to unreliable material behaviour. In order to investigate the structure-function relationships in CT covalent networks, monosodium 5-sulfoisophthalate (PhS) was selected as heparin-mimicking, growth factor-binding crosslinking segment, whilst 1,4-phenylenediacetic acid (4Ph) and poly(ethylene glycol) bis(carboxymethyl) ether (PEG) were employed as sulfonic acid-free diacids of low and high crosslinker length respectively. Hydrogels based on short crosslinkers (PhS and 4Ph) displayed increased crosslink density, decreased swelling ratio as well as minimal hydrolytic and lysozymic degradation, whilst addition of lysozymes to PEG-based networks resulted in 70 wt.-% mass loss. PhS-crosslinked CT hydrogels displayed the highest loss (40 ± 6 CFU%) of antibacterial activity upon incubation with *Porphyromonas gingivalis*, whilst respective extracts were tolerated by L929 mouse fibroblasts.

---

* Corresponding author. *Email address*: g.tronci@leeds.ac.uk (G. Tronci).

# 1. Introduction

Chitosan (CT) is a linear polysaccharide derived from the partial deacetylation of chitin, which is found in the exoskeletons of crustaceans and arthropods; as such, CT availability is second only to cellulose. The CT backbone consists of randomly repeating units of glucosamine and N-acetylglucosamine, whereby the former (i) mediate the solubility of CT in aqueous systems [1]; (ii) are amenable to chemical functionalisation towards the synthesis of water-soluble CT derivatives [2]; and (iii) provide CT with unique antibacterial properties [3]. With the breakout of antibiotic resistance, CT has found wide applications as haemostat [4], wound dressing [5] and antibacterial coating [6], in the form of electrospun webs [7], nano-particulates [8] or water-swollen networks [9]. As a nitrogen-rich polysaccharide, CT is insoluble in organic solvents, so that either extreme experimental conditions [10], oxidative degradation [11], multi-step synthetic routes [12], or combination thereof, are typically required to obtain customised material formats. These strategies lead to low yields and time-consuming purification, whilst native CT functionalities (e.g. degradability, antibacterial activity) may be irreversibly lost in resulting materials, resulting in poorly-controllable structure-function relationships. In an effort to synthesise functional systems in reliable, straightforward and mild environments, we have previously reported the formation of covalent hydrogel networks, whereby CT was reacted with various diacids in acid-free and room temperature conditions [13]. In light of the functionalisation imparted at the molecular level, these materials could be employed as tunable drug carriers of either negatively- or positively-charged soluble molecules, overcoming loading limitations linked to CT cationic charge.

Building on this knowledge, this study investigated the structure-function relationships in CT hydrogels functionalised with or without sulfonic acid pendant groups. The molecular scale of covalent networks was explored, whilst swelling properties in both acidic and basic pH, as well as hydrolytic and lysozymic degradability, quantified. Ultimately, the hydrogel's impact on both *Porphyromonas gingivalis* and L929 fibroblasts was also explored, looking at potential applicability of these materials in oral care.

## 2. Materials and methods

### 2.1 Materials

1-(3-Dimethylaminopropyl)-3-ethylcarbodiimide hydrochloride (EDC), N-hydroxysuccinimide (NHS) and Phosphate-buffered saline (PBS) solution were supplied by Alfa Aesar, VWR International and Lonza, respectively. All the other chemicals were purchased from Sigma.

### 2.2 Synthesis and characterisation of CT hydrogels

CT hydrogels were synthesised following previous protocols [13] (Supporting Information (SI)). The molar content of primary amino groups was measured via 2,4,6-trinitrobenzenesulfonic acid (TNBS) assay (n=2) [14, 15]. Respective degree of CT crosslinking (*C*) was quantified as:

$$C = \left(1 - \frac{Abs_{XL}}{Abs_{CT}}\right) \times 100 \qquad \text{(Equation 1)}$$

where $Abs_{XL}$ and $Abs_{CT}$ indicate the 343 nm-absorbances (UV-Vis, 6305 Jenway) of crosslinked and native CT, respectively.

### 2.3 Swelling and degradation tests

Weight-based swelling ratio (*SR*) was determined via 24-hour room temperature incubation of dry samples (~ 30 mg, n=3) in 5 mL 0.15 M NaCl aqueous solution at pH 4 – 9 [13]. Hydrolytic (5 ml PBS, pH 7.4, 37 °C) and lysozymic (5 ml PBS, pH 7.4, 37 °C, 2 mg·ml$^{-1}$ lysozyme) degradation tests were performed for 30 days. At specific times, samples (n=3) were collected, rinsed with distilled water, paper-blotted, weighed, and freeze-dried. Time point-specific relative mass ($\mu_{rel}$) was determined as:

$$\mu_{rel} = \left(1 - \frac{m_t}{m_d}\right) \times 100 \qquad \text{(Equation 2)}$$

where $m_t$ and $m_d$ correspond to the time point-specific and initial dry masses, respectively. Time point-specific *SRs* were determined from respective swollen and dry samples masses.

### 2.4 Cytotoxicity and antibacterial tests *in vitro*

L929 cells were cultured on hydrogel extracts (section S4, SI). Cell morphology and metabolic activity were analysed via transmitted light microscopy (Zeiss, Germany) and MTS assay (CellTiter 96® AQueous Assay, Promega), respectively.

*Porphyromonas gingivalis* W50 (*P. gingivalis*) were cultured on UV-irradiated samples (Bop-Rad GS Gene Linker[TM] UV, 250 mJ, 3x). Biofilms were grown on each sample (n=3) and anaerobically incubated in brain heart infusion (BHI) broth (80% $N_2$, 10% $CO_2$, 10% $H_2$ atmosphere, 37 °C, 48 h) (Whitley A45 Anaerobic Workstation, Don Whitley Scientific Ltd). Each sample was then transferred into 2 ml sterile reduced transport fluid (RTF) and vortexed. 1 ml $10^{-4}$ dilution in sterile RTF was spread unto blood agar plates (Columbia blood agar base, Oxoid). The plates were anaerobically incubated (37 °C, 48 h) and number of colonies optically counted (Stuart SC6).

## 3. Results and discussion

Network formation was accomplished via a nucleophilic addition/elimination mechanism of CT primary amino groups with NHS-activated diacids (Scheme S1, SI). An averaged degree of crosslinking was measured via TNBS calorimetric assay in the range of 8 – 26 mol.-%; samples CT-4Ph and CT-PEG displayed the highest and lowest crosslink densities, respectively (Table 1), in agreement with previous studies [1, 10, 14]. TNBS values of PhS-crosslinked CT were used to determine the molar content of sulfonic acid moieties, considering a 1:2 molar ratio between the former and PhS-crosslinked amino groups. An averaged molar content of 13 mol.-% PhS was observed, in agreement with $^1$H-NMR measurements (18 mol.-% PhS) on degraded network oligomers [13].

**Table 1.** Degree of crosslinking (*C*) and swelling ratio (*SR*) of chitosan networks obtained via either TNBS assay or equilibration in aqueous solutions at varied pH, respectively. *SR* data could not be determined for native CT controls in light of their non-controllable degradation behaviour in aqueous environments.

| Sample ID | *C* /mol.-% | *SR* /wt.-% | | | |
|---|---|---|---|---|---|
| | | pH 4 | pH 6.5 | pH 7.4 | pH 9 |
| CT-PhS | 26 ± 1 | 542 ± 80 | 498 ± 254 | 717 ± 168 | 582 ± 247 |
| CT-4Ph | 38 ± 1 | 409 ± 55 | 548 ± 103 | 472 ± 27 | 527 ± 65 |
| CT-PEG | 8 ± 4 | 1090 ± 86[(*)] | 1098 ± 188[(*)] | 1195 ± 142[(*)] | 1133 ± 173[(*)] |

[(*)] Statistically significant mean at each selected pH ($p <0.05$, Bonferroni test)

Introduced sulfonic acid groups in network CT-PhS could be exploited as a simple heparin-mimicking cue for binding and delivering growth factors [6] towards mineralised as well as soft tissue repair in dentistry. Their impact on hydrogel swelling was investigated in acidic and basic pH, as descriptive of native and bacteria-contaminated tissue states, respectively [16]. *SR* values were in the range of 409 ± 55 – 1195 ± 142 wt.-%, proving to be lower than the hydrogels prepared with lower CT concentrations [5]. Samples CT-PhS displayed a slightly increased *SR* in neutral compared to acidic solutions, in contrast to the other hydrogels, which is attributed to the deprotonation of both PhS sulfonic acid ($pK_a \sim 3$) and CT primary amino ($pK_a \sim 6.5$) groups [14]. Observed pH-dependent swelling behaviour may be exploited to design CT-based passive sensors for bacteria detection. Overall, samples of CT-PEG displayed the highest *SR*, due to their low degree of crosslinking and the hydrophilicity of PEG (Table 1), showing promise as a dressing material for moist wound healing.

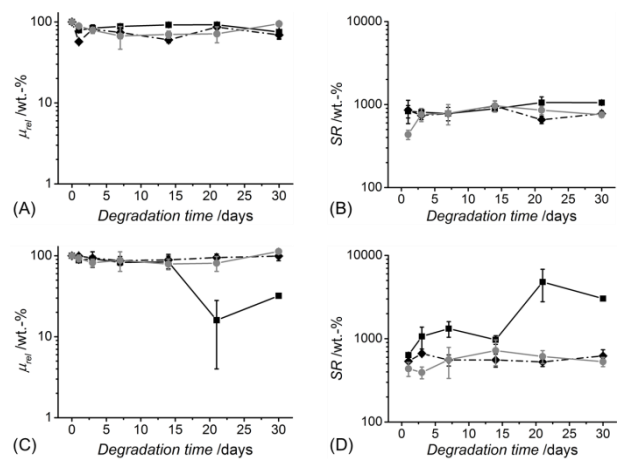

**Figure 1.** Relative mass ($\mu_{rel}$) and swelling ratio (*SR*) profiles of networks CT-PhS (—●—), CT-4Ph (—·—♦—·—) and CT-PEG (–■–) during hydrolytic ((A-B) and lysozymic (C-D) degradation. Native CT controls are not reported as completely degraded after 1-day incubation in both testing media.

Minimal changes in sample weight and swelling ratio were recorded during hydrolytic incubation, whilst native CT samples were completely degraded in either hydrolytic or lysozymic media, as expected for a non-crosslinked material. Addition of lysozymes proved to induce nearly 70 wt.-% mass loss in hydrogels CT-PEG (Figure 1), suggesting that cleavage of CT chains was preferentially accomplished via lysozyme-induced breakdown of glycosidic linkages rather than hydrolysis of amide net-points. The increased degradation kinetics of hydrogels CT-PEG was in agreement with their low degree of crosslinking. SEM analysis revealed an increased porous sample microstructure following 30-d lysozymic incubation (Figure 2), suggesting a surface erosion rather than bulk degradation mechanism.

Figure 3 describes the impact of hydrogels *in vitro* on both L929 mouse fibroblasts and *Porphyromonas gingivalis*, respectively. Cell morphology cultured on extract of sample CT-PhS was comparable to the ones of cells cultured on DMEM. The high tolerability of hydrogel extract was also confirmed by cellular metabolic activities. Hydrogels CT-PhS cultured with *Porphyromonas gingivalis* also displayed the lowest retention of native CT antibacterial activity, followed by hydrogels CT-4Ph (56 CFU%) and CT-PEG (69 CFU%) (Figure 3, D).

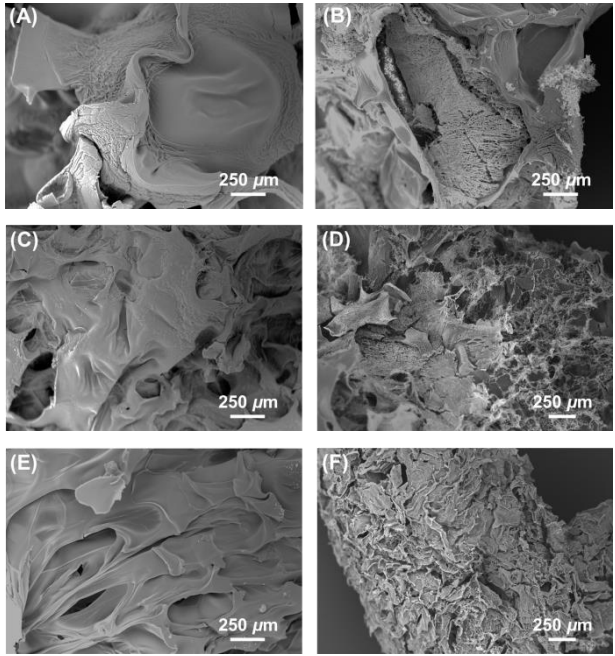

**Figure 2.** Scanning electron microscopy of surface structure following either network formation (A, C, E) or 30-d enzymatic incubation (B, D, F). (A-B): CT-PhS; (C-D): CT-4Ph; (E-F): CT-PEG.

The electrostatic interaction of positively-charged amino groups with negatively-charged microbial cell membranes has been hypothesised to be responsible for the antibacterial activity of CT [3]. Results obtained in this study support this mechanism: in light of the consumption of primary amino groups and introduction of negatively-charged sulfonic acid moieties, a reduced antibacterial activity in samples CT-PhS is expected. The lower antibacterial activity in samples CT-Ph compared to CT-PEG is in line with the lower molar content of primary amino groups in the former, compared to the latter, formulation.

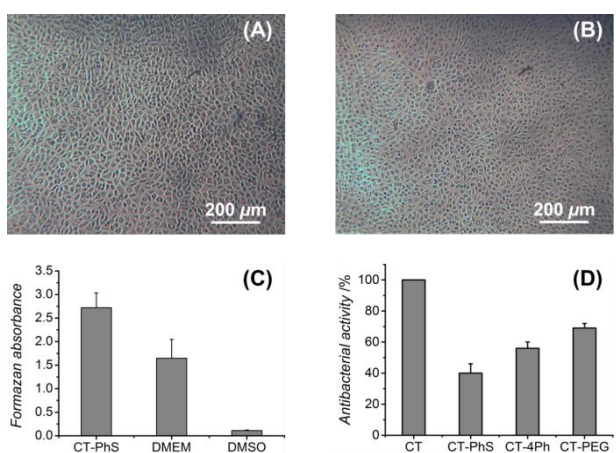

**Figure 3.** (A-B): extract cytotoxicity tests were conducted according to EN DIN ISO standard 10993-5 and L929 cell morphology investigated following 48-hour culture in either DMEM (A) or extract of sample CT-PhS (B). (C): MTS assay following 48-hour cell culture on either extract of sample CT-PhS, DMEM or DMSO. (D): antibacterial activity displayed by both CT hydrogels and native CT following 48-hour incubation with *P. ging*.

## 4. Conclusions

CT hydrogels were synthesised with either a diacid bearing heparin-mimicking sulfonic acid group or two sulfonic acid-free diacids. Hydrogel degradability occurred preferentially via lysozyme-induced breakdown of glycosidic linkages, whereby crosslink density played a significant role. Sulfonic acid pendant groups proved to induce a significantly reduced antibacterial activity in hydrogels CT-PhS, whilst no hydrogel-triggered toxic response was observed with L929 fibroblasts.

**Acknowledgements**

The authors wish to thank Assoc. Prof. Hiroharu Ajiro (NAIST, JP) and Prof. Mitsuru Akashi (Osaka University, JP) for the initial scientific discussions. The financial support from the Clothworkers' Centre for Textile Materials Innovation for Healthcare is gratefully acknowledged.

**Supporting information**

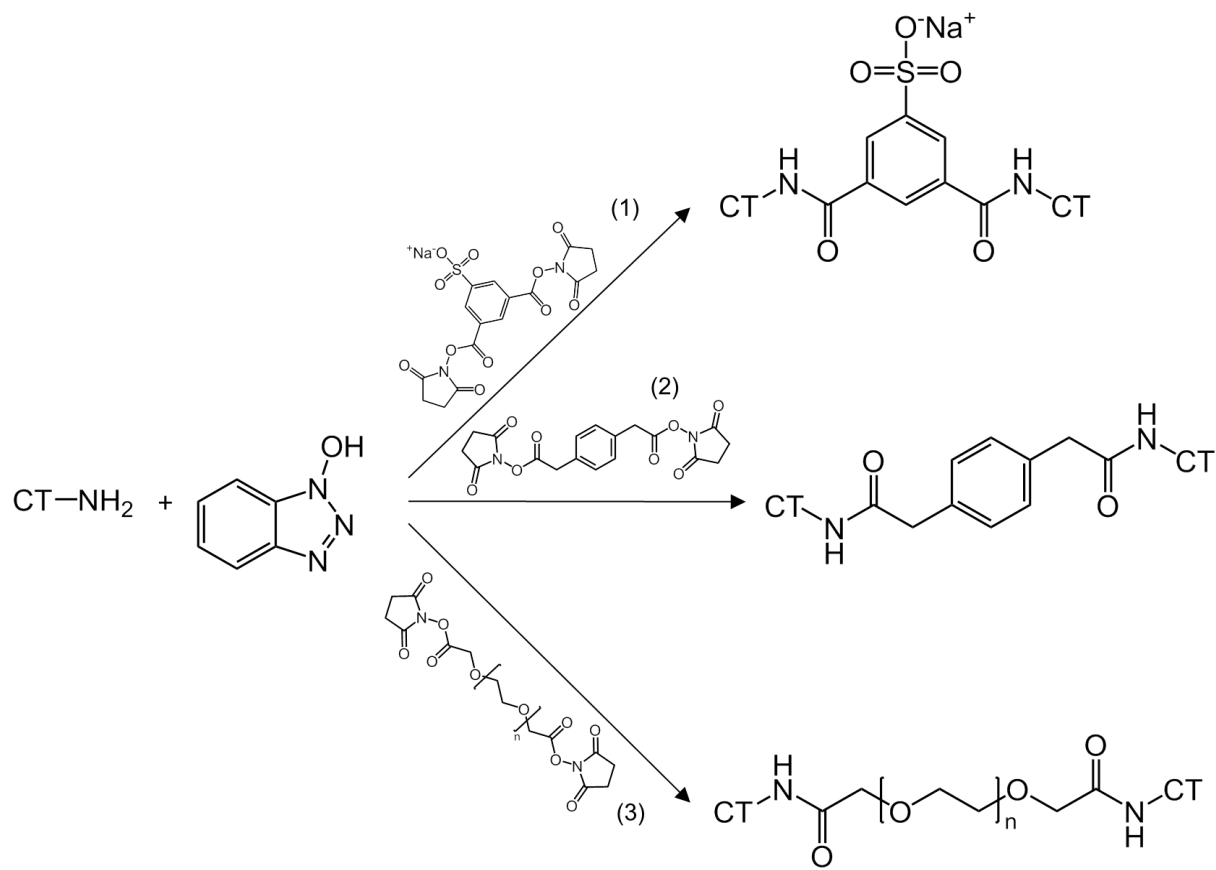

**Scheme S1.** Formation of CT hydrogels: CT was dissolved in a HOBt-H$_2$O system and reacted with NHS-activated PhS (1), Ph (2) and PEG (3), respectively.

**S1. Synthesis of CT hydrogels**

CT (0.15 g) was dissolved in a distilled water solution (4.23 g) containing HOBt (0.12 g) via magnetic stirring at room temperature. A 2-fold molar content of either PhS, 4Ph or PEG (with respect to the amino groups in the CT solution) was activated in distilled water (0 °C, 30 min.) with a 6-fold molar ratio of both EDC (2.3 M) and NHS (2.3 M) with respect to crosslinker carboxylic functions. NHS-activated solution was mixed with previously-obtained CT solution and incubated overnight at room temperature under gentle shaking. Formed hydrogels were washed with distilled water and freeze-dried. Control samples were prepared by freeze-drying HOBt-containing aqueous solution of CT. Crosslinked samples are coded as CT-XXX, whereby 'XXX' identifies the selected diacid.

**S2. Scanning electron microscopy**

The internal architecture of dry freshly synthesised and partially degraded samples was inspected using a Hitachi SU8230 FESEM. Samples were inspected with a beam intensity of 10 kV after gold sputtering using a JFC-1200 fine sputter coater.

**S3. Statistical analysis**

Data normality was confirmed via the Shapiro-Wilk test and data presented as average ± standard deviation. One-way ANOVA was carried out to test statistical significance ($p < 0.05$, Bonferroni test).

**S4. Preparation of hydrogel extracts for cytotoxicity assay**

0.1 mg of ethanol-treated hydrogel was incubated at 37 °C in 1 mL Dulbecco's Modified Eagle Medium ((DMEM). After 72 hours, sample extract (n=3) was applied to 80% confluent L929 mouse fibroblasts cultured on a polystyrene 96-well plate. Dimethyl sulfoxide (DMSO) and DMEM were used as negative and positive controls, respectively (n=3).

**References**

[1] Matsumoto M, Udomsinprasert W, Laengee P, Honsawek S, Patarakul K, Chirachanchai S. A Water-Based Chitosan–Maleimide Precursor for Bioconjugation: An Example of a Rapid Pathway for an In Situ Injectable Adhesive Gel. Macromol. Rapid Commun. 2016, DOI: 10.1002/marc.201600257.

[2] Schiffman JD, Schauer CL. Cross-Linking Chitosan Nanofibers. Biomacromolecules 2007; 8; 594-601.

[3] Perelshtein I, Ruderman E, Perkas N, Tzanov T, Beddow J, Joyce E et al. Chitosan and chitosan–ZnO-based complex nanoparticles: formation, characterization, and antibacterial activity. J. Mater. Chem. B 2013; 1; 1968 - 1976.

[4] Dowling MB, Kumar R, Keibler MA, Hess JR, Bochicchio GV, Raghavan SR. A self-assembling hydrophobically modified chitosan capable of reversible hemostatic action. Biomaterials 2011; 32; 3351-3357.

[5] Kumar PTS, Lakshmanan V-K, Anilkumar TV, Ramya C, Reshmi P, Unnikrishnan AG, et al. Flexible and Microporous Chitosan Hydrogel/Nano ZnO Composite Bandages for Wound Dressing: In Vitro and In Vivo Evaluation. ACS Appl. Mater. Interfaces 2012; 4; 2618-2629.

[6] Almodóvar J, Mower J, Banerjee A, Sarkar AK, Ehrhart NP, Kipper MJ. Chitosan-Heparin Polyelectrolyte Multilayers on Cortical Bone: Periosteum-Mimetic, Cytophilic, Antibacterial Coatings. Biotechnology and Bioengineering 2013; 110; 609-618.

[7] Salihu G, Goswami P, Russell S. Hybrid electrospun nonwovens from chitosan/cellulose acetate. Cellulose 2012, 19, 739-749.

[8] Birch NP, Schiffman JD. Characterization of Self-Assembled Polyelectrolyte Complex Nanoparticles Formed from Chitosan and Pectin. Langmuir 2014; 30; 3441-3447.

[9] Tachaboonyakiat W, Serizawa T, Akashi M. Hydroxyapatite Formation on/in Biodegradable Chitosan Hydrogels by an Alternate Soaking Process. Polymer Journal 2001; 33; 177-181.

[10] Ciulik CB, Bernardinelli OD, Balogh DT, de Azevedo ER, Akcelrud L. Internal plasticization of chitosan with oligo(DL-lactic acid) branches. Polymer 2014; 55; 2645-2651.

[11] Lu Y, Slomberg DL, Schoenfisch MH. Nitric oxide-releasing chitosan oligosaccharides as antibacterial agents. Biomaterials 2014; 35; 1716-1724.

[12] Jou C-H, Yang M-C, Suen M-C, Yen C-K, Hung C-C, Hwang M-C. Preparation of O-diallylammonium chitosan with antibacterial activity and cytocompatibility. Polym Int 2013; 62; 507-514.

[13] Tronci G, Ajiro H, Russell SJ, Wood DJ, Akashi M. Tunable Drug-loading Capability of Chitosan Hydrogels with Varied Network Architectures. Acta Biomater 2014; 10; 821-830.

[14] Malhotra M, Tomaro-Duchesneau C, Saha S, Kahouli I, Prakash S. Development and characterization of chitosan-PEG-TAT nanoparticles for the intracellular delivery of siRNA. International Journal of Nanomedicine 2013; 8; 2041-2052.

[15] Tronci G, Doyle A, Russell SJ, Wood DJ. Structure-property-function relationships in triple helical collagen hydrogels. Mater. Res. Soc. Symp. Proc. 2013; 1498; 145-150.

[16] Schneider LA, Korber A, Grabbe S, Dissemond J. Influence of pH on wound-healing: a new perspective for wound-therapy? Arch Dermatol Res. 2007; 298; 413-420.